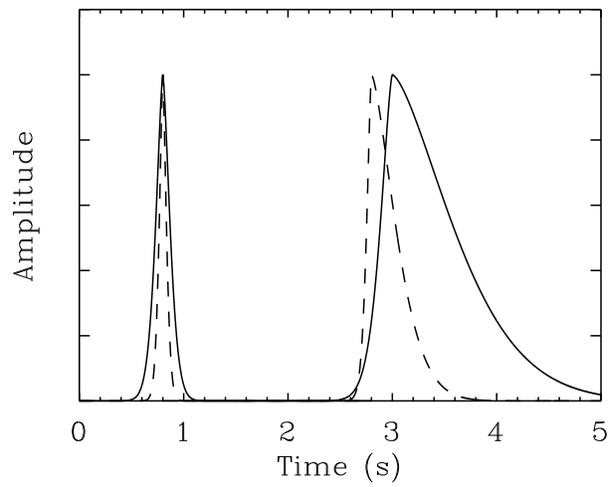

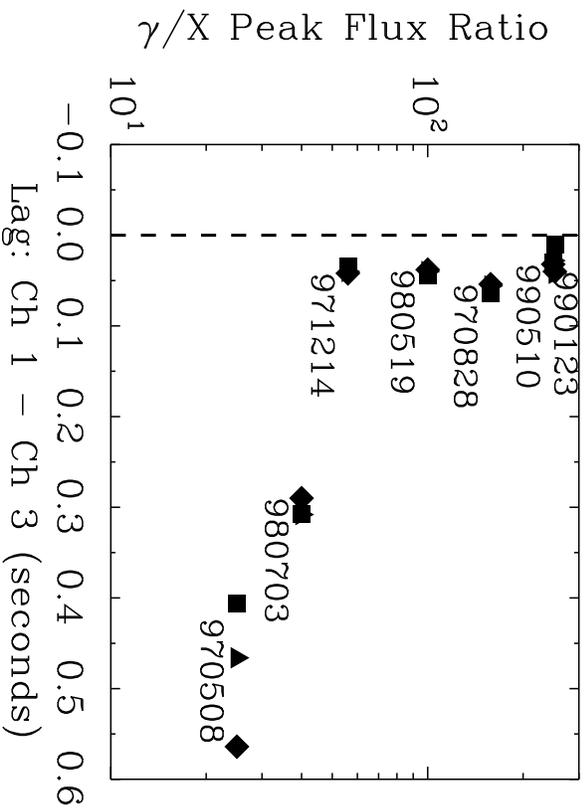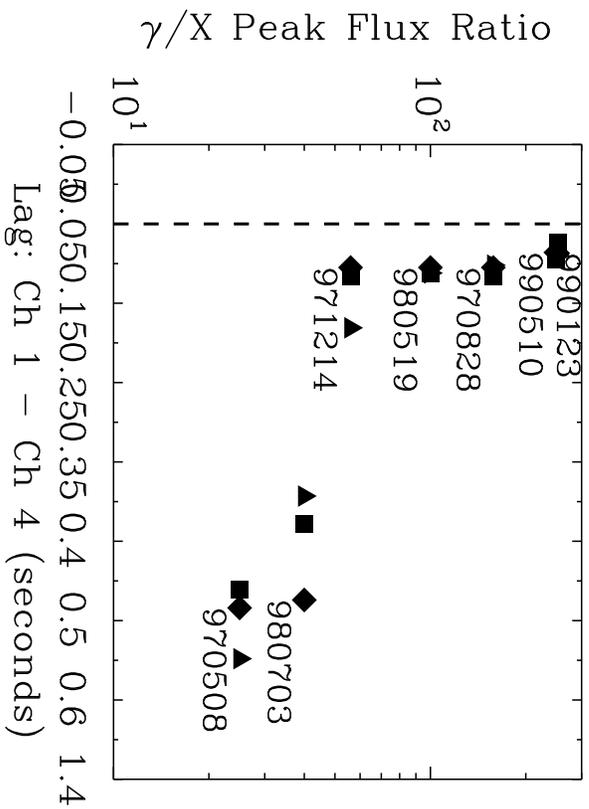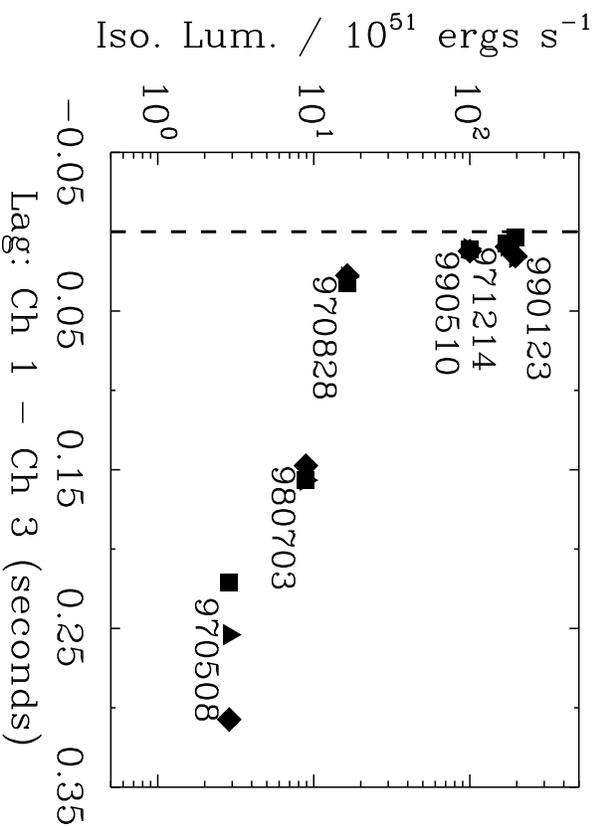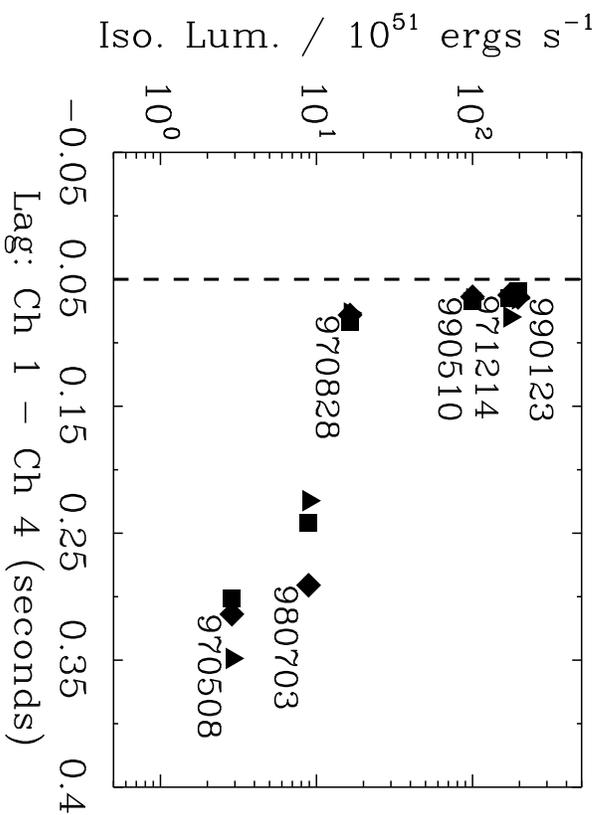

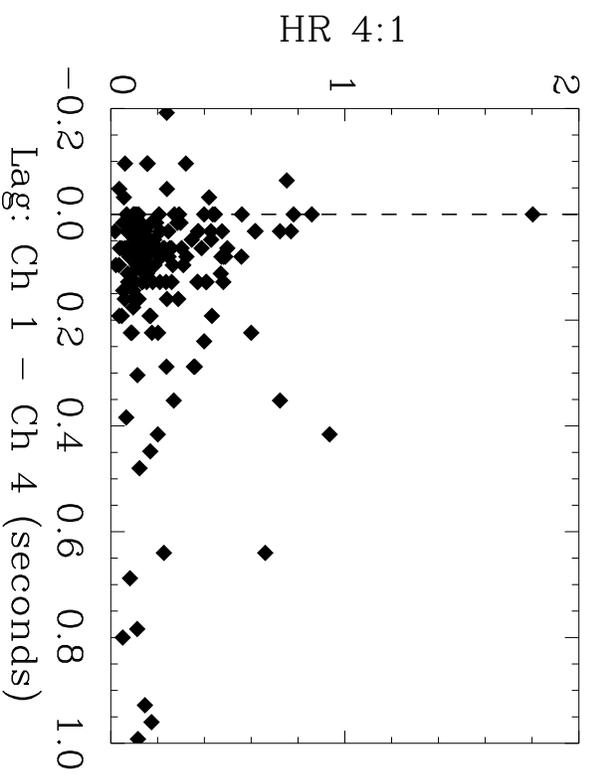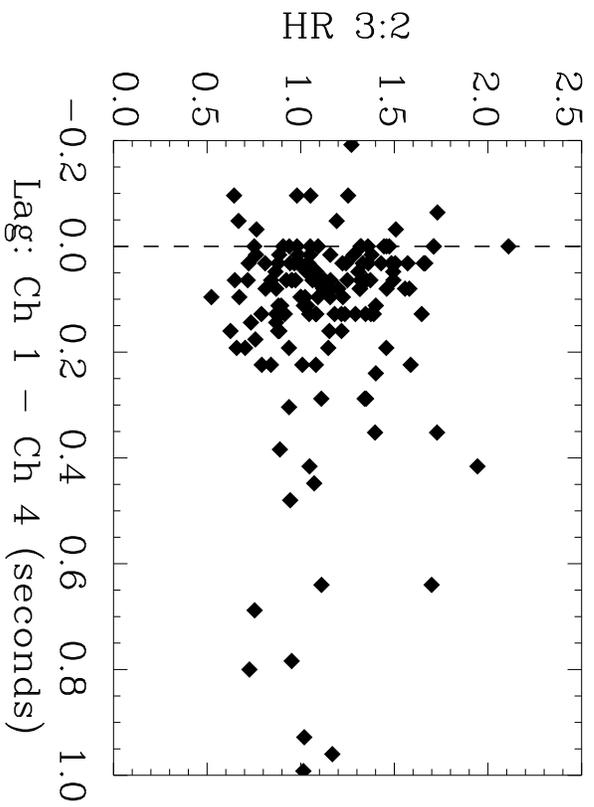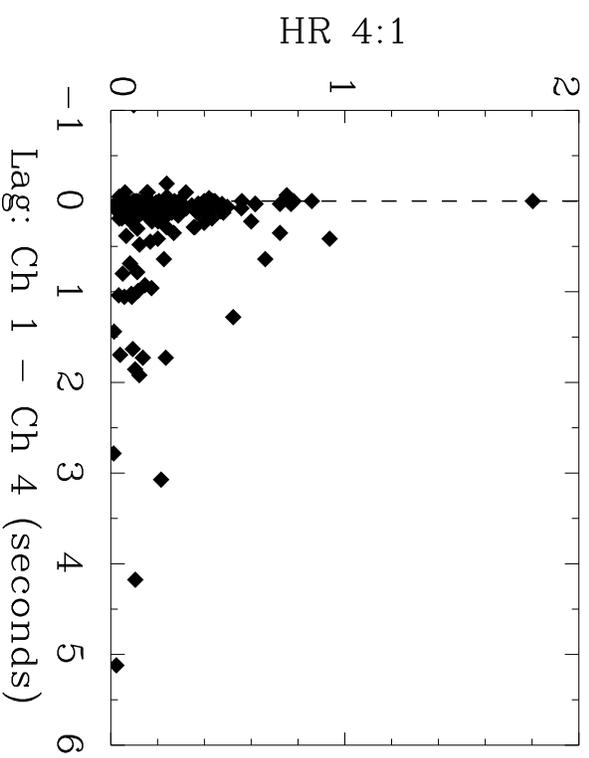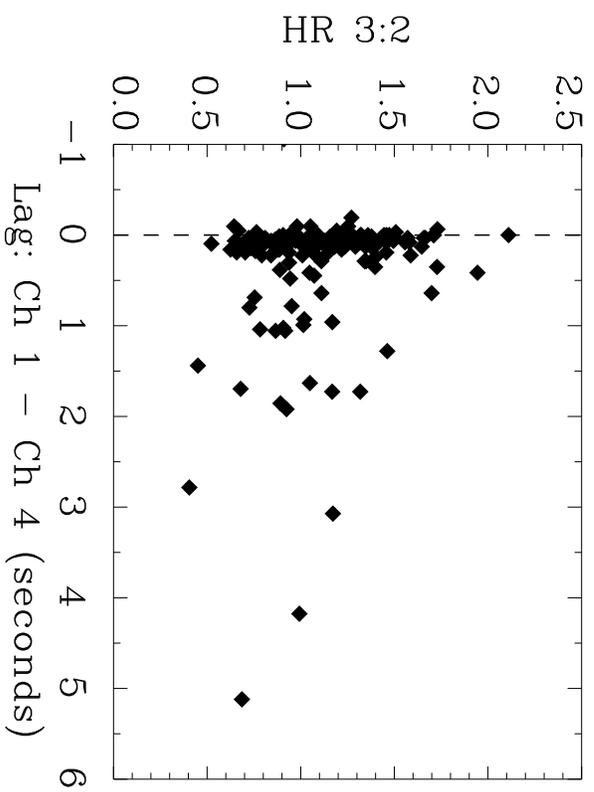

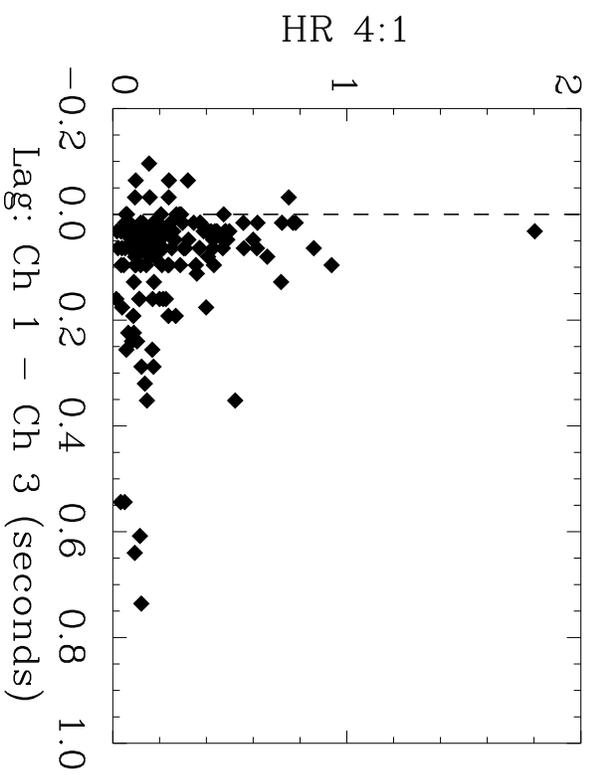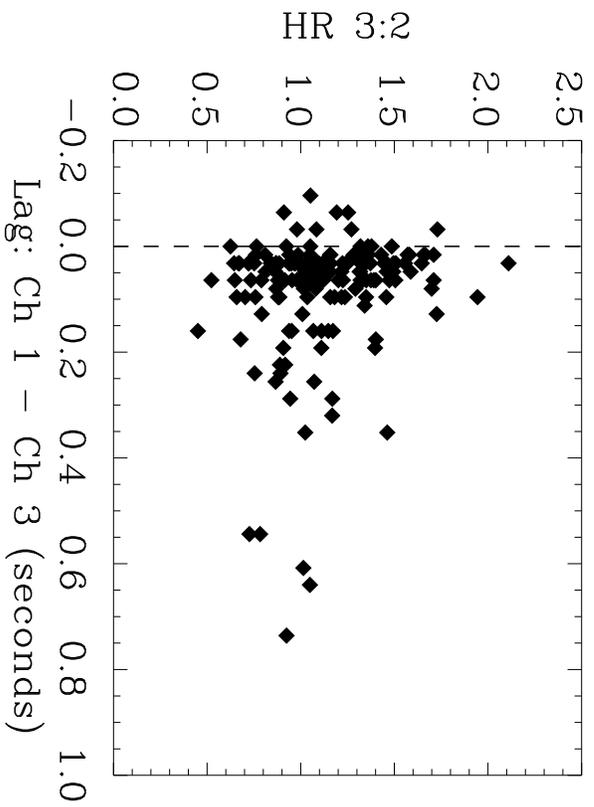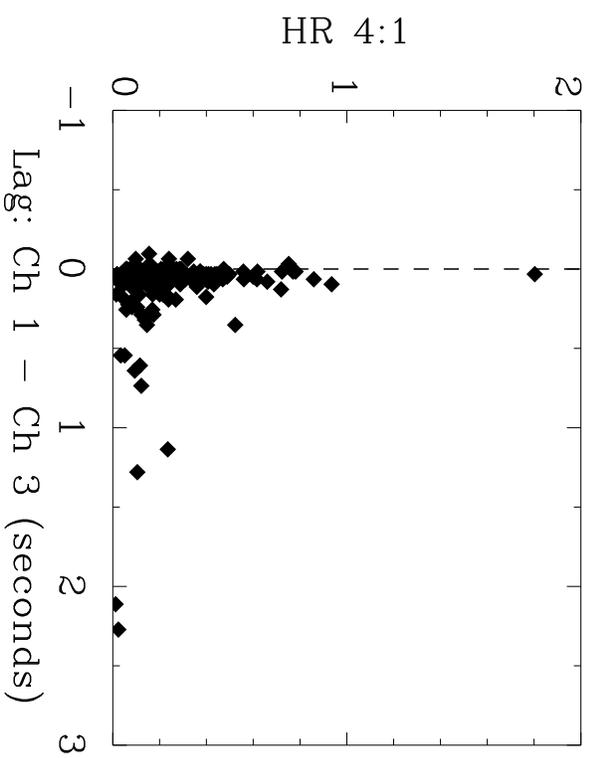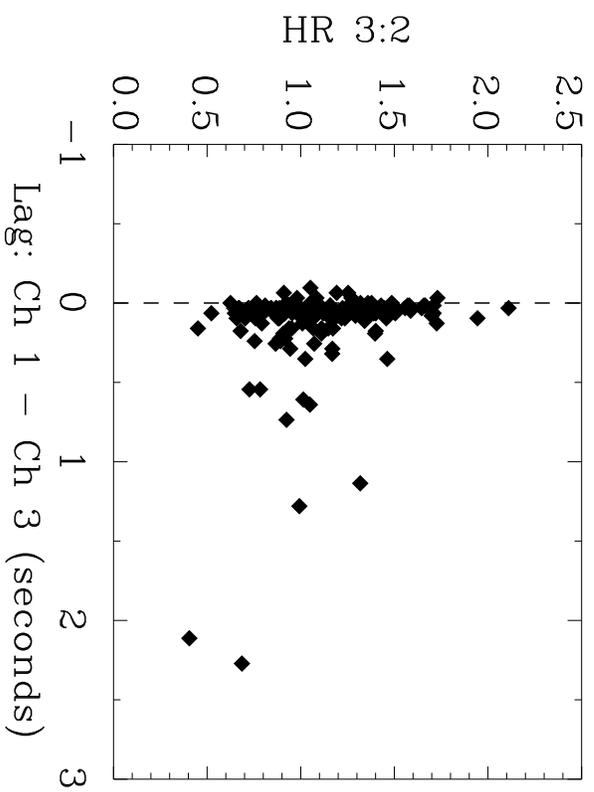

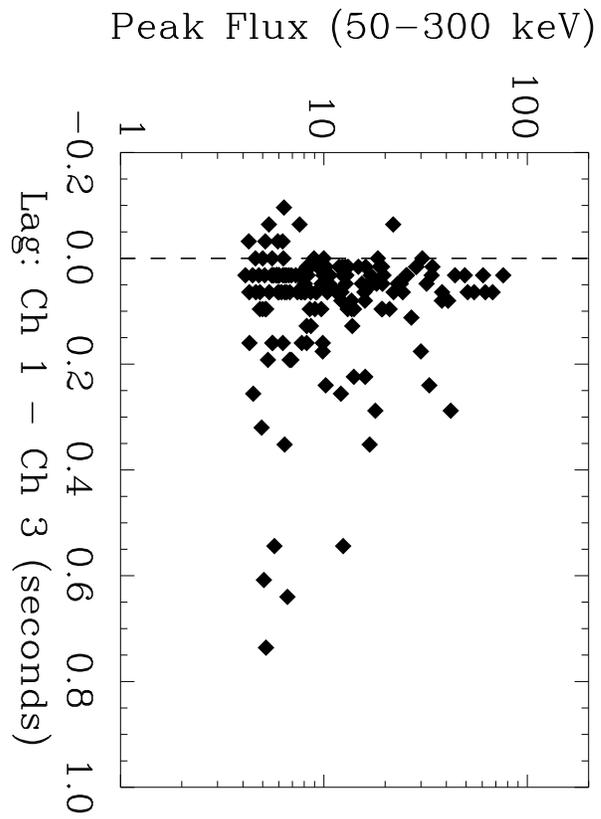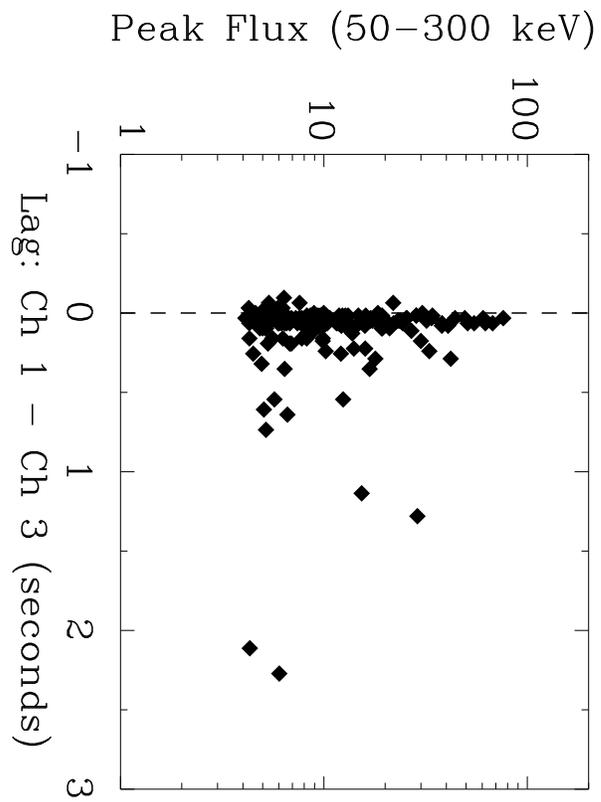

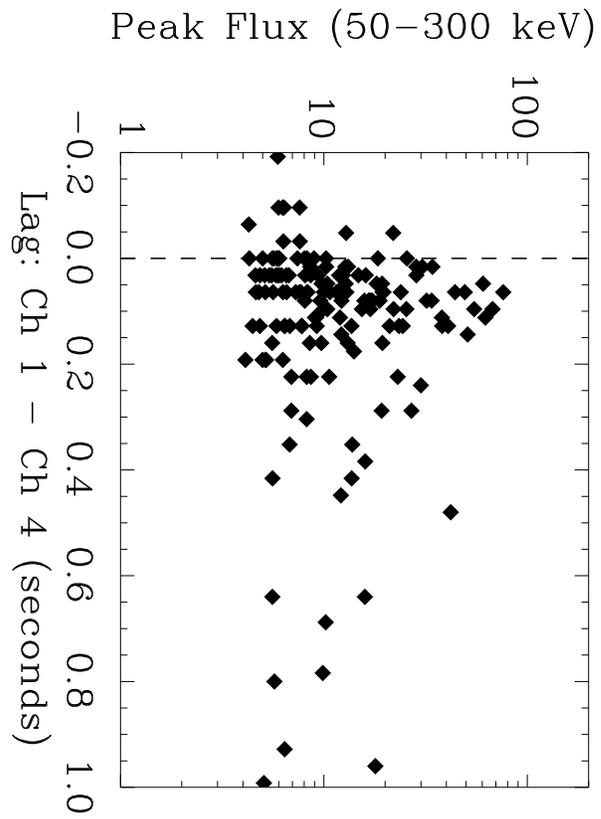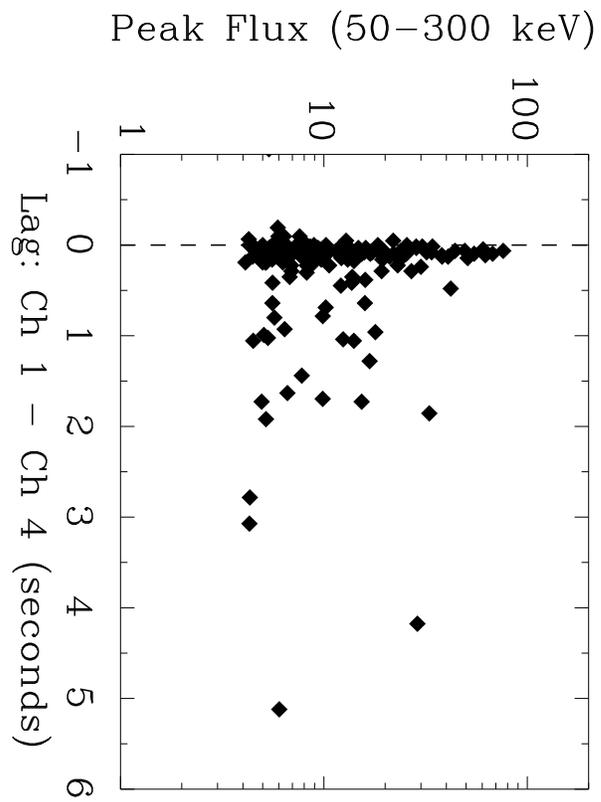

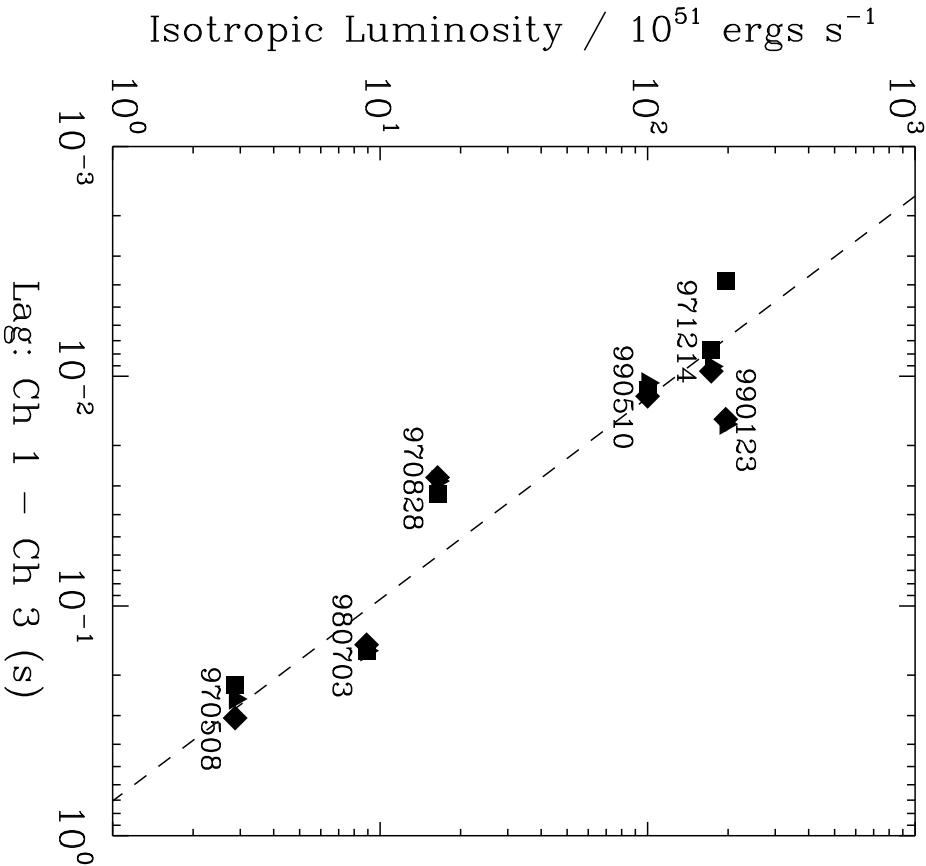
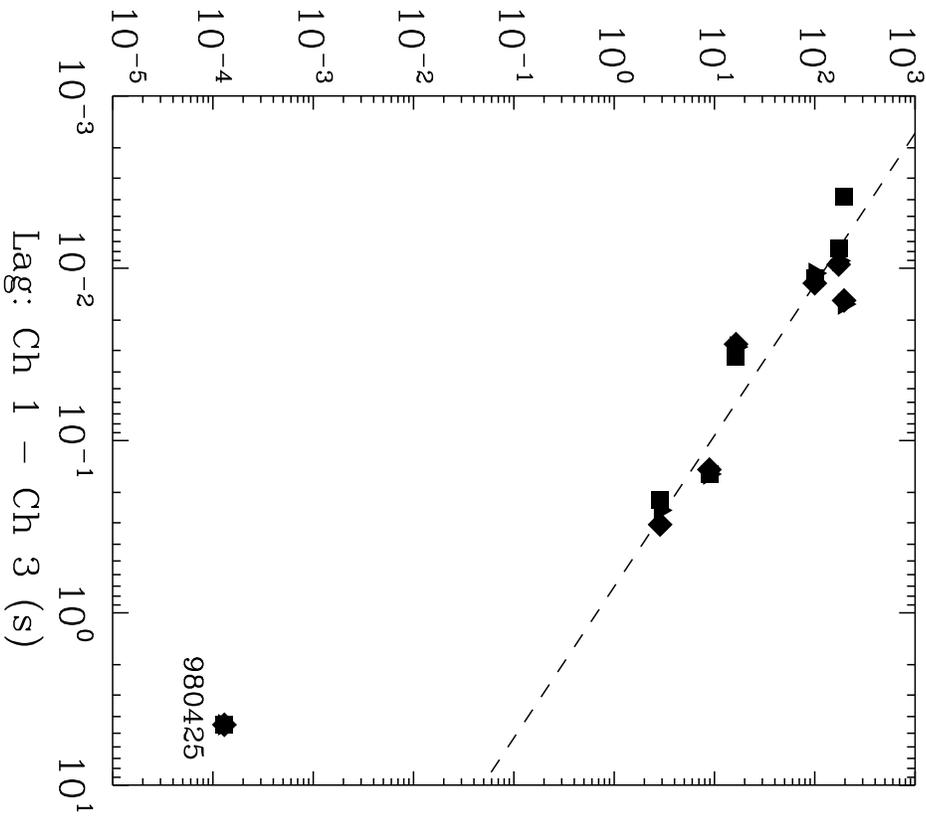

# CONNECTION BETWEEN ENERGY-DEPENDENT LAGS AND PEAK LUMINOSITY IN GAMMA-RAY BURSTS

J. P. Norris,[1] G. F. Marani[1,2], and J. T. Bonnell,[1,3]

[1] NASA/Goddard Space Flight Center, Greenbelt, MD 20771.

[2] National Research Council

[3] Universities Space Research Association.

# ABSTRACT

We suggest a connection between the pulse paradigm at gamma-ray energies and the recently demonstrated luminosity distribution in gamma-ray bursts: The spectral evolution timescale of pulse structures is anticorrelated with peak luminosity, and with quantities which might be expected to reflect the bulk relativistic Lorentz factor, such as spectral hardness ratio. We establish this relationship in two important burst samples using the cross-correlation lags between low (25–50 keV) and high (100–300 keV and > 300 keV) energy bands. For a set of seven bursts (six with redshifts) observed by CGRO/BATSE and BeppoSAX which also have optical or radio counterparts, the γ/X peak flux ratios and peak luminosities are anti-correlated with spectral lag. For the 174 brightest BATSE bursts with durations longer than 2 s and significant emission above 300 keV, a similar anti-correlation is evident between gamma-ray hardness ratio or peak flux, and spectral lag. For the six bursts with redshifts, the connection between peak luminosity and spectral lag is well-fitted by a power-law, $L_{53} \approx 1.3 \times (\tau/0.01 \text{ s})^{-1.15}$. While GRB 980425 (if associated with SN 1998bw) would appear to extend this trend qualitatively, with a lag of ~ 4–5 s and luminosity of ~ $1.3 \times 10^{47}$ ergs s$^{-1}$, it falls below the power-law relationship by a factor of several hundred. As noted previously by Band, most lags are concentrated on the short end of the lag distribution, near 100 ms, suggesting that the GRB luminosity distribution is peaked on its high end.

Subject headings: gamma rays: bursts – temporal analysis

# 1. INTRODUCTION

The phenomena of spectral evolution in gamma-ray bursts (GRBs) are two distinct, observed effects: Pulse peaks migrate to later times and become wider at lower energies. And, burst spectra tend to soften as the event progresses, such that the individual, evolving pulses have impressed upon them an envelope which governs a global spectral decay. Evidence for pulse spectral evolution was first described in Norris (1983), based essentially on the behavior of two pulses, and corroborated in analysis of several bursts observed by SMM (Norris et al. 1985). Global spectral evolution was suggested (Norris et al. 1987) and demonstrated as a general – but not universal – trend in analyses of large BATSE burst samples. Band (1997) performed a cross-correlation analysis between 4-channel data to demonstrate the trend of spectral softening on the timescale of pulses in an integral sense. Using the cross-correlation method, Cheng et al. (1995) found that soft emission had a time delay relative to high-energy emission. Ford et al. (1995) examined higher resolution spectra of bright bursts, confirming the tendency of bursts to soften as they progress. This global spectral behavior is arguably related to burst temporal asymmetry on long timescales (Link, Epstein, & Priedhorsky 1993; Nemiroff et al. 1994). Progressively longer decays are now detected from hard X-ray energies (Connaughton 1999) to wavelengths stretching into the radio bands, and must be manifestations of radiation transfer, and of dissipation of the burst energy preceding and during the afterglow phase.

The physical explanation for pulse evolution is probably more closely connected to the primal energy generation, rather than dissipation, mechanism. From analysis of many individual pulses in bright, long BATSE bursts, Norris et al. (1996) proposed that a "pulse paradigm" operates in GRBs: Among pulses where shape is relatively well determined, the rise-to-decay ratio is unity or less; as this ratio decreases, pulses tend to be wider, the pulse centroid is shifted to later times at lower energies, and pulses tend to be spectrally softer. The width/asymmetry aspect of this paradigm is reproduced in Figure 1 (and appears to apply to pulses in short bursts; see Figure 2,

Norris 1995). The correlation of pulse spectral hardness with pulse symmetry was less clear; Norris et al. remarked that convolution with the global trend of spectral softening may dilute the correlation of individual pulse hardness with symmetry within a sample of bursts. Also, from visual inspection, many investigators have observed that pulses within a given burst tend to exhibit a large degree of self-similarity, or little variation in pulse width. In particular, Fenimore, Ramirez-Ruiz, & Wu (1999) quantitatively demonstrated via autocorrelation analysis that different intervals within GRB 990123 exhibit comparable pulse widths. Ramirez-Ruiz & Fenimore (1999) arrived at the same conclusion for bright bursts in general via two analysis approaches. Their kinematic arguments based on constancy of pulse width provide conclusive proof that GRB pulses can be generated only by a small central engine, otherwise the deceleration expected in external shock models would be evidenced by widening pulses as the burst progresses (see additional arguments against external shocks: Fenimore, Madras, & Nayakshin 1996; Sari & Piran 1997; Fenimore et al. 1999; Fenimore & Ramirez-Ruiz 1999).

However, for lack of a Rosetta Stone, the physical import of the pulse paradigm has remained undeciphered. This situation appears changed by the detection of several optical counterparts to GRBs, which have disparate implied isotropic luminosities and total energies. The detection and observations of GRB 990123 across the spectrum (Akerlof et al. 1999; Bloom et al. 1999; Connors et al. 1999; Djorgovski et al. 1999a; Feroci et al. 1999; Kippen 1999) indicate that this burst would hold the current record for gamma-ray energy release, $> 3 \times 10^{54}$ ergs for isotropic emission, except that the temporal decay of the optical afterglow exhibits a break after ~ 3 days, from a power-law index of $\approx 1.1$ to a steeper decline (Kulkarni et al. 1999) – the signature predicted for a burst originally beamed into a small solid angle, and later undergoing transverse expansion (Rhoads 1997; 1999). The high energy and Lorentz factor inferred for GRB 990123 beamed into a small solid angle ($\theta \sim 0.1$) would therefore be more commensurate with that inferred for other GRBs with optical counterparts and associated redshifts where clear evidence for strong beaming is lacking, ~ $10^{52}$ ergs (Kulkarni et al.). From the work of Fenimore and colleagues, one might expect concomitant differences to prevail near the primal cause, and therefore some obvious manifestation of different gamma-ray spectral/temporal signatures. (See especially, the elucidation of correspondence between active portion of emitting shell and wide versus narrow pulse structures in Fenimore, Madras, & Nayakshin [1996] and Fenimore & Ramirez-Ruiz [1999].)

In this work, we analyze cross-correlation lags between energy bands for two important GRB samples: seven events detected by BeppoSAX and BATSE, six of which have associated redshifts; and the 174 brightest BATSE bursts with significant flux > 300 keV. For the bursts with associated redshifts, lag is anti-correlated with $\gamma/X$ peak flux ratio and isotropic peak luminosity, following a power-law relationship. For the large sample we show that there exists an anti-correlation between energy-dependent lag, and both BATSE spectral hardness ratio and peak flux. We mention central questions attending these results: SN 1998bw and GRB 980425; multiplicity of pulses and extreme spectral lags; complications in deriving a GRB luminosity distribution from the spectral lag information, due to time-dilation and redshift effects; and expectations for observing optical afterglows from short GRBs.

## 2. CROSS-CORRELATION ANALYSIS

We first considered all GRBs detected by BeppoSAX with associated optical afterglows (in one case, with a radio transient) and with usable BATSE data through May 1999. This sample includes seven bursts: GRBs 970508, 970828, 971214, 980519, 980703, 990123, and 990510. Other GRBs with detected optical counterparts were not usable in this analysis: the necessary BATSE data types for GRB 980326 and 980613 were incomplete or not recorded; GRB 980329 was eliminated due to corruption of MER data in high channels that affected DISCSC data during intense portions of the burst. The first burst for which X-ray and optical afterglows were detected, GRB 970228, was not seen by BATSE; however, from published BeppoSAX data we are able to estimate upper limits on a spectral lag. Table 1 describes the adopted sample of GRBs, including gamma-ray and X-ray peak fluxes, redshift determinations, and inferred isotropic luminosities. This sample is particularly relevant since some indication of source distance has been reported, or has been or may be constrained in each case from the observations at longer wavelengths. In the case of GRB 970828, no optical counterpart was observed, but a GRB-like radio transient consistent with the BeppoSAX X-ray position was detected after the burst (Frail & Kulkarni 1997), and found to be coincident with a dust lane within a galaxy of redshift z = 0.958 (Djorgovski & Odewahn 1997; S. Djorgovski, private communication). Also listed in Table 1 is GRB 980425, possibly associated with SN 1998bw. The redshift and luminosity quoted in Table 1 assume this still controversial association is correct. Cross-correlation results for this burst are discussed in section 4.

The larger sample is defined by the following requirements: $T_{90}$ duration > 2 s; BATSE peak flux (50–300 keV) > 4 photons cm$^{-2}$ s$^{-1}$; and peak count rate (> 25 keV) > 14000 count s$^{-1}$. The design of the latter two requirements was an automated first attempt to include bursts with significant emission in BATSE channel 4 (> 300 keV) of the Large Area Detectors (LADs) so that we might compute spectral lags across the broadest energy range with high time resolution BATSE data. Still, from visual inspection of the 4-channel LAD profiles, we eliminated sixteen bursts which had (nearly) imperceptible emission in channel 4. We also eliminated one burst, trigger # 5614 – the brightest burst observed by BATSE – whose peak rates resulted in counter overflow during the most intense pulse. The resulting bright sample contained 174 bursts. Not only do these bursts afford a usable signal in channel 4, but they are also less inhomogeneous than the full BATSE sample in terms of time dilation and redshift effects.

The 4-channel time profiles were constructed following procedures described in Bonnell et al. (1997), which require timing overlap for the 1.024-s continuous DISCLA data and the 0.64-ms PREB and DISCSC data types. The background fitting and $T_{90}$ duration estimation procedures are also described in Bonnell et al. Backgrounds were fitted separately for each of the four energy channels (25–50, 50–100, 100–300, > 300 keV), and subtracted to yield signal profiles. For GRB 970828 the DISCSC data were not complete; we therefore used the 16-channel MER data binned to approximate the four DISCSC channels.

For many bursts, pulse fitting is problematic, due to overlapping structures: such decomposition methods – invaluable for characterizing tractable regions – can yield ambiguous results in crowded regions. Fortunately, as discussed in the introduction, Nature appears to have produced in GRBs a phenomenon in which pulse width is often essentially constant during the gamma-ray portion of the event. As discussed in Fenimore, Ramirez-Ruiz, & Wu (1999), this implies that the bulk Lorentz factor, $\Gamma$, is also nearly constant. Thus if there are relationships between luminosity, $\Gamma$, and spectral lag, we expect their expressions to be preserved by integral measures, such as cross correlation of time profiles in two different energy channels.

We therefore computed spectral lags using a cross correlation method similar to that described in Band (1997), but differing only in apodization details: For the sample of eight bursts in Table 1, we cross correlated the signal in channel 1 with those in channel 3 and (for seven bursts with suitable data) channel 4, including regions down to fractions 0.1, 0.3, and 0.5 of the peak intensity (> 25 keV). Performing the analysis across three different intervals allows us to address the question of evolution of cross-correlation function (CCF) lag. We limited the analysis to regions with count rates $> 0.1 \times$ peak intensity for two reasons: to minimize inclusion of post nonburst emission – GRB tails can continue much longer than the more prompt onsets – and to minimize overlap with possible hard X-ray afterglow, which may manifest spectral behavior different than that of the GRB itself. To compute the CCF between signals in two channels over a finite range, each time series must be concatenated with apodization intervals as long as the desired maximum lag, prior to and after the burst. Since the intervals analyzed do not extend to background levels, we assigned constant values to the apodization intervals, computed as the intensity averages of the respective 10 adjacent time bins, per channel, for the pre and post burst intervals. If there were some relationship between lag and $\Gamma$, then for constant $\Gamma$, one would expect the CCFs for the three intensity regions to yield identical lags. However, the signal in channel 4 can be negligible during some portions of the burst, and so long intervals with essentially noise contribute to the CCF. The result is a less precise measurement of the CCF peak due to random walk. Thus for the optical counterpart bursts, some of which are relatively dim, we deemed it necessary to compute CCFs for the three intensity regions and compare results. The problem is practically nonexistent for channel 3 since the burst signal persists nearly as long as that in channel 1. For the sample of 174 bright bursts, the channel-4 signal is usually adequate, so we computed CCFs for only the region down to 0.1 of peak intensity. Also, it would be difficult to illustrate three lag determinations for such a large number of bursts on one plot in a straightforward manner.

In a perfect experiment, the data would have much finer resolution than the temporal feature sizes, high signal-to-noise (s/n) levels would result in smooth CCF curves, and side lobes in the CCF would always be lower in amplitude than the central lobe – altogether yielding, automatically without inspection, the clear CCF lag peak. In reality, it is sometimes expeditious to bin the time series more coarsely to achieve reasonable s/n levels, and it is advisable to examine simultaneously on multiple passes the lagged time series, noting the response of the CCF to peculiarities of each time series correlation. Also, in a noisy series the lag is often more accurately measured by fitting the region near the CCF main peak, rather than by simply recording the lag of the peak itself.

After multiple inspections, we decided to compute CCFs for the large sample of bright bursts with peak intensities $\lesssim$ 29500 counts s$^{-1}$ at 128 ms (82 events); brighter bursts were analyzed at 64 ms resolution (74 events). Initially, all CCFs were fitted using a range of 9 bins (irrespective of temporal resolution) using a quadratic form. For 40 bursts, the residuals between the fit and CCFs for channel 4 to 1 (CCF41), or channel 3 to 1 (CCF31), were too large, either near the peak or the extrema, and so the fit range was lowered to 7 bins for these bursts. The fit was iterated, reassigning the central bin of the fitted range to the CCF maximum, until the determination of peak time converged. Also, we found that a cubic form better represented the tendency towards asymmetric CCFs. Although the higher order yielded only a secondary refinement – since the fitting range had already been made sufficiently narrow to avoid inadequate fits – a cubic was deemed necessary in some cases to obtain good residuals; hence for the large sample we utilized a cubic globally to minimize manual intervention.

For the optical counterpart sample we carefully studied the dependence of lag measure on degree of polynomial fit and on the range of CCF fitted. A primary motivation was to determine if the lags for different intensity ranges were significantly different. We therefore estimated statistical errors for the CCFs via the bootstrap approach, creating 101 realizations for each burst and each intensity range: To the original time profiles (before background subtraction), we added Poisson-distributed noise, per energy channel. The peak time was determined by over-resolving the time dimension of the model fit by a factor of 32 (rather than 4 for the large sample), since we required sufficient precision to determine the $16^{th}$ ($-1\ \sigma$) and $84^{th}$ ($+1\ \sigma$) ranked measurements for the bootstrap errors. Both quadratic and cubic forms were fitted to the CCF31 curves; only the quadratic was fitted for CCF41, since the s/n levels for the channel 4 profiles do not support the additional degree of freedom. We observed each of the fits from a set of $101 \times 8 \times 3 \times 2$ realizations, being sure that (1) the fitted range was sufficiently long so that fits were always concave down, but (2) short enough so that peak and extrema were well fitted. Range adjustments were made where necessary and the process was iterated several times. Generally, the bursts with larger lags required the longest CCF range to be fitted to realize the first condition, whereas bursts with smaller lags required shorter CCF ranges to fulfill the second condition. However, GRB 971214 (with small lag) was dim, and required a relatively longer fit range to realize stable, concave down fits. Table 2 lists the CCF fitted ranges, median ($50^{th}$ rank) lag determinations with plus and minus one sigma errors (*without* corrections for time dilation), for the seven GRBs and three intensity ranges.

The ranges having been thus optimized, the lag measurement is then basically dependent on polynomial degree. In Table 2 note that for CCF31 the cubic fits generally yielded smaller lags with larger fractional errors than did the quadratic fits (cf. Wu & Fenimore 1999). Cubic fits are more obliging with the additional degree of freedom, and they probably represent the true lag more accurately: Near its peak the CCF for bursts is almost invariably asymmetric with the steeper slope on the zero lag side (the asymmetry reflects ubiquitous low energy lag). Therefore the CCF is better fitted by a cubic, whose peak will move to accommodate the peak and asymmetric wings, evidencing the variance across the several realizations. The peak of the quadratic form, intrinsically symmetric, will tend to lie at larger lag values, since in the fitting process the two quadratic branches will split the difference of the CCF's asymmetry. A concomitant effect is that the fractional error for the quadratic fit will tend to be smaller since it cannot accommodate asymmetry. We therefore chose to utilize the CCF31 cubic fits for most of the analysis described below.

## 3. RESULTS

Figure 2 illustrates the measured lags for CCF41 (top) and CCF31 (bottom) plotted versus the γ/X peak flux ratio (left) for the seven bursts in Table 1 (excluding GRB 980425), and versus isotropic peak luminosity (right) for the subset of six bursts with associated redshifts. The symbols denote lags determined for included signal intensities to 0.1 (diamonds), 0.3 (triangles), and 0.5 (squares) × peak intensity (> 25 keV). Generally, the CCF41 lags are ~ 2–3 × longer than the CCF31 lags. For a given burst all three intensity ranges usually yield results within the statistical errors, with the spread in median values ranging up to ~ 15–20%. The fractional errors for CCF31 lags concentrate in the range ~ 20–40%. For the right two panels (ordinate is peak luminosity), the lags are converted to the presumed frame of the burster object, multiplying by $(1+z)^{-1}$ to remove cosmic time dilation.

The one significant exception to the trend for lag agreement between intensity ranges is GRB 990123. The lags for intensity ranges 0.1 and 0.3 are smaller than for intensity range 0.5, by all three fit measures listed in Table 2: ~ 3–4 × smaller for CCF31, and ~ 2 × smaller for CCF41. This is a robust result, in that varying the fitted CCF range does not make the difference disappear. A possible hint of the same trend is seen in Figure 2 for GRB 970508, for both the CCF31 and CCF41 values, but in neither case is the difference significance.

A trend is evident for bursts with larger γ/X ratio or higher luminosity to have shorter lags. The two bursts with the longest lags, GRBs 970508 (z = 0.835) and 980703 (z = 0.967), were relatively nearer, but dimmer, bursts. Their inferred isotropic gamma-ray peak luminosities were correspondingly smaller, $3\times10^{51}$ and $9\times10^{51}$ ergs s$^{-1}$, respectively, compared to the luminosities inferred for GRBs 971214 (z = 3.412), 990123 (z = 1.600) and 990510 (z=1.619): $173\times10^{51}$, $196\times10^{51}$, and $100\times10^{51}$ ergs s$^{-1}$, respectively, all of which have short lags. Deconvolution of the effect of redshift on lag for this sample should result in monotonic corrections to first order. We will attempt this larger task in subsequent work using BATSE MER data with 16-channel energy resolution.

For the sample of 174 bright BATSE bursts, only HRs and peak fluxes are available. Thus the interpretation of the lag determinations is complicated by several smearing factors: the lag coordinate contains a cosmological time-dilation factor and the concomitant spectral redshift, which modifies the counts in the lagged energy channels as well as the ordinate value, be it HR or peak flux; and the peak flux contains the unknown distance. Nevertheless, we may expect to see a residual signature if a general anti-correlation between luminosity and lag exists, for two reasons. First, these bursts represent the brightest one-seventh of the triggered bursts in the BATSE sample of long bursts. Hence, while there is a range of a factor of twenty in peak flux, some of which must represent intrinsic variation in luminosity, the range in measured average time dilation is small (Marani et al. 1999). Second, HR is modified only by redshift. Therefore we examined both types of relationship, HR versus CCF lag, and peak flux versus CCF lag. Here, HR is defined as the ratio of total counts, in the region above $0.1 \times$ peak intensity (>25 keV), between two LAD channels.

Figure 3 illustrates the CCF41 lags plotted versus HR3/2 (top) and HR4/1 (bottom) for the 174-burst sample. On the left and right sides the lag coordinate extends to 1 s and 6 s, respectively; the left plots magnify the region near the origin, while the right plots include bursts with large lags and relatively low HRs. The trend towards an occupied triangular region is evident. In all four plots, the point with the highest HR is GRB 990123. Similarly, Figure 4 illustrates the CCF31 lags plotted versus the same HRs. The lags are shorter, hence on the left and right sides of the plot the coordinate extends to 1 s and 3 s, respectively. The same trend is evident as in Figure 3. Generally, bursts with longer CCF lags tend to have lower hardness ratios. As previously noted by Band (1997), the CCF lags are concentrated near the origin, predominantly populating the region < 200 ms (< 100 ms) for CCF41 (CCF31).

A similar trend is evident for CCF lag plotted versus peak flux, as shown in Figure 5. The CCF41 (CCF31) lags on top (bottom) are plotted versus BATSE peak flux (50–300 keV) with compressed (extended) lag range on the left (right) plots. The fact that the anti-correlation trend is obvious separately for peak flux and hardness ratio suggests that the different extrinsic effects (including distance for peak flux) convolved in these ordinate measures do not completely obscure a more fundamental relationship between spectral lag and luminosity.

## 4. DISCUSSION

To recapitulate, previously the pulse paradigm reported in GRBs (Norris et al. 1996) described the tendency for wide (narrow) pulses to be more (less) asymmetric, to peak later at lower energy (nearly simultaneously) and to be spectrally softer (harder). As noted by many investigators, degree of asymmetry and pulse width often appear to hold fairly constant with a given burst. Overall spectral evolution within a burst makes quantification of the spectral aspect of the paradigm somewhat difficult. However, temporal analyses of burst structure by two methods, confirming the tendency of self-similarity within a burst, allow the interpretation that the Lorentz factor does not change substantially and suggest that integral measures of lag across substantial intervals of a burst can provide acceptable diagnostics (Fenimore, Ramirez-Ruiz, & Wu 1999; Ramirez-Ruiz & Fenimore 1999).

In this work we demonstrated for the sample of six bursts with associated redshifts, that isotropic peak luminosity appears to be related to spectral lag, as measured by the cross-correlation of low and high energy LAD channels in the BATSE data. The recent burst, GRB 990123, has the highest inferred isotropic peak luminosity among bursts with optical counterparts, and one of the smallest spectral lags found in this study when corrected for cosmic time dilation (~ 0.006–0.02 s). The two GRBs with lowest peak luminosities, 970508 and 980703, have the longest lags (~ 0.1–0.2 s). We note that GRB 970228 (z = 0.695, Djorgovski 1999b) is intermediate in peak luminosity (~ $5\times10^{51}$ erg $s^{-1}$) between these lower luminosity bursts. Although lack of BATSE data and differing energy ranges preclude a direct comparison of GRB 970228 with our sample, from visual inspection of Figure 1 in Frontera et al. (1998), we estimate a lag between the 2–10 keV and 40–700 keV bands of ~ 2 s. This measure may best be calibrated using Figure 2 of Piro et al. (1999) for GRB 970508, where we estimate the lag between the 1.5–26 keV and 40–700 keV bands to be ~ 2–3 s, taking into consideration counting statistics. Both estimates were made by considering the lags between peaks of the initial pulses, which were the most intense in each case. Thus for GRB 970228, a guess for the lag between our 25–50 keV and 100–300 keV bands – which span a much tighter range than the BeppoSAX energy bands – including correction for time dilation, might be ~ 100–300 ms, similar to that for GRB 970508.

Similar relationships between spectral lag and peak flux or spectral hardness ratio are evident in the sample of 174 bright bursts. However, these measurements have unknown extrinsic factors – time dilation, redshift, and distance – which must obscure any clearer intrinsic relationship, and which represent considerably larger systematic uncertainties than inherent in the statistical errors for the spectral lag determinations. For example, in our bright burst sample the dynamic range in peak flux is a factor of 20 (for the triggered BATSE sample, almost a factor of $10^3$), across which the majority of lags are distributed with minima of ~ 50–100 ms. Given the presently observed range of GRB luminosity distances, 2.6–32 Gpc (including GRB 990712), the occurrence of short lags ranging over a factor of 20 in peak flux may be attributable to distance alone.

The spectral lag / luminosity relationship for our subset of six bursts with known redshifts is better appreciated in a log-log plot. Figure 6a, similar to the lower right panel of Figure 2, plots log[CCF31] versus log[$L_{51}$]. The dashed line is a power-law fit to the lags for intervals including count rates > 0.1 × peak intensity (diamond symbols). The fit yields

$$L_{53} \approx 1.3 \times (\tau / 0.01 \text{ s})^{-1.14}, \qquad (1)$$

where $L_{53}$ is the luminosity in units of $10^{53}$ ergs s$^{-1}$ and $\tau$ is the CCF31 lag (cubic fit). Power-law fits to lags for the 0.3 and 0.5 intensity ranges yield indices of 1.15 and 1.03, respectively. The CCF31 points for GRB 990123 exhibit the largest range on the log-log plot. As mentioned above, the differences appear to be significant (see Table 2), but the relatively coarse 64 ms binning of DISCSC data is a consideration. The actual measured values before correction for time dilation are 0.040, 0.042, and 0.010 s for intervals with > 0.1, 0.3, and 0.5 × peak intensity, respectively; thus we are attempting measurements which overresolve the single-sample resolution of the DISCSC data. The corresponding values for the CCF41 lags in GRB 990123 are relatively more closely spaced: 0.074, 0.080, and 0.048 – all within a factor of ~ 1.7.

As far we understand, this would appear to be the first "luminosity—color" relationship for GRBs constructed from gamma-ray data, albeit calibrated using luminosities inferred from optical redshifts – similar to the HR diagram for main sequence stars, or the period-luminosity relationship for Cepheids. If corroborated with subsequent measurements, Equation (1) would be another rung in the cosmic distance-scale ladder: The expectation is that spectral lags for the general population of GRBs would translate into luminosities, once the lags are corrected for extrinsic effects.

In Figure 6b the luminosity and lag ranges are expanded to include GRB 980425 ($\equiv$ SN 1998bw ?), which consisted of a single long pulse (Norris, Bonnell, and Watanabe 1999). We estimated the spectral lag (~ 4.5 s) from the average of the difference of times for channels 1 and 3 at half maximum on the rise and decay, (~ 2 s and ~ 7 s, respectively); the difference in peak times for the same two channels is ~ 3 s. (A CCF41 lag cannot be had since the burst was not detected above 300 keV.) GRB 980425, while exhibiting a qualitatively similar signature – relatively long spectral lag, and very low luminosity – falls below the fitted power-law relationship for the six bursts at cosmological distances by a factor of ~ 400–700, depending on the precise powe-law index adopted. We conclude that either GRB 980425 is truly a different kind of GRB – "on a different branch of the GRB HR diagram" – or else it is not associated with SN 1998bw.

As noted previously by Band (1997), and confirmed in this work, most lags are concentrated on the short end (CCF41 $\lesssim$ 200 ms; CCF31 $\lesssim$ 100 ms). And, most bright bursts, for which it is straightforward to make the observation, have complex structure, with many narrow pulses; few bright bursts can be characterized as one- or two-pulse, long smooth events with long lags (see Fenimore et al. 1999). Thus, if lag is indeed generally related to luminosity, this would suggest that the luminosity distribution is peaked more towards its high, rather than low, end. Note that bursts 10 × fainter than those used in this study actually triggered BATSE. Accurate CCF analysis would require more painstaking effort due to lower S/N levels, but important differences might be found. We intend to pursue a larger program, investigating dimmer bursts and the possibility of deconvolving extrinsic effects from the spectral lags, to determine if a unique relationship between lag and luminosity can be constructed for the general sample. One might envision an iterative approach using data with higher spectral resolution to obtain estimates of deredshifted spectral lags: there are approximately two free parameters (isotropic luminosity and redshift), and even with only 4-channel data seven integral determinants are available (three lags plus four average fluxes). Note however, that producing accurate redshifts directly from gamma-ray data would appear to be tricky: a trial redshift moves narrower pulses from higher energy into the analysis band; narrower pulses have smaller lags, which would dictate a higher redshift trial in the next iteration. Regardless, in order to construct a GRB Hubble plot – physical luminosity versus redshift – the redshifts will need to be determined from longer wavelength observations.

There are related questions for which direct investigative approaches may be difficult. Short bursts ($T_{90} < 2$ s) obey the asymmetry/width part of the pulse paradigm (Norris 1995) but the connection between spectral hardness and lag is yet to be investigated. Short bursts have pulses compressed by a factor of 10–20 compared to pulses in long bursts, and have slightly higher hardness ratios (Kouveliotou 1993). Does this imply that short bursts have higher luminosities? Short bursts do not appear to manifest extended, low-intensity hard X-ray tails (Connaughton 1999); this may not bode well for frequent detection of optical decays and redshift determinations. However, if a similar connection between spectral lag and luminosity could be established from a few optical afterglows (or galaxian hosts coincident with X-ray counterparts) for short bursts, then the distance scale for short bursts might be established.

It is possible that the anti-correlations found between the energy-dependent lag and γ/X peak flux ratio, BATSE spectral hardness ratios, and isotropic γ-ray peak luminosity also contain indirect connections to the Lorentz factor $\Gamma$. The γ/X peak flux ratio during a burst expresses the spectral hardness ratio across the widest available range at high energy and thus would be related to $\Gamma$ if the peak in ν•F(ν) depends on $\Gamma$. The γ-ray luminosity may be related to $\Gamma$ as well: Suppose that the primal cause of GRBs is particle acceleration in strong electric fields created by rapidly rotating black holes, with the field strength proportional to angular velocity. Then higher $\Gamma$'s would be associated with harder spectra and higher luminosities.

The explication of observed energy-dependent lags in current GRB models employing internal shocks is inconclusive. Panaitescu & Mészáros (1998), using hydrodynamic simulations and including effects from the geometrical curvature of the emitting source, do not predict any lag. On the other hand, Daigne & Mochkovitch (1998) do predict lags in their model of internal shocks from a relativistic wind, although they did not take into account any geometrical effects, and the lags are too large for most bursts: ~ 5 s between ~ 25–50 keV and > 300 keV. While Sari & Piran (1999) argue that time profiles cannot provide any information on the initial value of the Lorentz factor in the internal shock scenario, Fenimore, Ramirez-Ruiz, & Wu (1999) conclude that the GRB source is a central engine but it may not necessarily be powered by internal shocks.

In conclusion our most useful result is that spectral lag appears to be anticorrelated with peak luminosity in GRBs with associated redshifts, following an approximate power law with slope $\alpha \approx -1.1$. If this relationship is confirmed with a larger sample, the import is evident: Calibration of spectral lags should lead to estimates of physical luminosities and thus a Hubble plot for GRBs.

We thank the anonymous referee for several incisive and substantive suggestions.

REFERENCES

Akerlof, C.W., et al.  1999, Nature, submitted

Band, D. L.  1997, ApJ, 486, 928

Bloom, J. S., et al.  1998, ApJ, 508, L21

Bloom, J. S., et al.  1999, ApJ, submitted, astro-ph/9902182

Bonnell, J. T., Norris, J.P., Nemiroff, R.J.,  & Scargle, J.D.  1997, ApJ, 490, 79

Cheng, L. X., et al.  1995, A&A, 300, 746

Connaughton, V.  1999, in preparation

Connors, A., et al.  1999, GCN Notice 230

Daigne,F., & Mochkovitch, R.  1998, MNRAS, 296, 275

Djorgovski, S. G., Odewahn, S.C.  1997, IAUC 6730

Djorgovski, S. G., et al.  1998a, GCN Notice 79

Djorgovski, S. G., et al.  1998b, ApJ, 508, L17

Djorgovski, S. G., et al.  1999a, GCN Notice 251

Djorgovski, S. G., et al.  1999b, GCN Notice 289

Fenimore, E. E., Madras, C.D., & Nayakshin, S.  1996, ApJ, 473, 998

Fenimore, E. E.,  et al.  1999, ApJ, 512, in press, astro-ph/9802200

Fenimore, E. E., Ramirez-Ruiz, E., & Wu, B.  1999, ApJ, submitted, astro-ph/9902007

Fenimore, E. E., & Ramirez-Ruiz, E.  1999, astro-ph/9906125

Feroci, M. et al.  1999, IAUC 7095

Ford, L. et al.  1995, ApJ, 439, 307

Frail, D.A., & Kulkarni, S.  1997, IAUC 6730

Fruchter, A. et al.  1999, astro-ph/9902236

Frontera, F. et al. 1998, ApJ, 493, L67

Galama, T. et al.  1998, ApJ, 497, L13

Heise, J. et al.  IAUC 6787

in't Zand, J., et al. 1998, ApJ, 505, L119

Kippen, R.M. 1999, GCN Notice 224

Kouveliotou, C., et al. 1993, ApJ, 413, L101

Kulkarni, S. R., et al. 1998, Nature, 393, 35

Kulkarni, S. R., et al. 1999, Nature, submitted, astro-ph/9902272

Levine, A. et al. 1998, IAUC 6966

Link, B., Epstein, R.I., & Priedhorsky, W.C. 1993, ApJ, 408, L81

Marani, G.F., et al. 1999, in preparation

Metzger, M. et al. 1997, IAUC 6655

Muller, J. et al. 1998, IAUC 691

Nemiroff, R.J., et al. 1994, ApJ, 423, 432

Norris, J. P. 1983, PhD thesis, University of Maryland, NASA TM 85031

Norris, J. P., et al. 1985, ApJ, 301, 213

Norris, J. P., et al. 1987, Adv. Space Res., 6, 19

Norris, J. P. 1995, Ap&SS, 231, 95

Norris, J. P., et al. 1996, ApJ, 459, 393

Norris, J.P., Bonnell, J.T., and Watanabe, K. 1999, ApJ, 518, 901

Piro, L. et al. 1997, A&A, 331, 41

Palazzi, E., et al. 1998, GCN Notice 48

Panaitescu, A., & Mészáros, P. 1998, ApJ, submitted, astro-ph/9810258

Ramirez-Ruiz, E., & Fenimore, E. E. 1999, A&A Supp., submitted, astro-ph/9812426

Rhoads, J.E. 1997, ApJ, 478, L1

Rhoads, J.E. 1999, preprint

Sari, R., & Piran, T. 1997, ApJ, 485, 270

Sari, R., & Piran, T. 1999, astro-ph/ 9901338

Waxman, E. 1997, ApJ, 489, L36

Wu, B. & Fenimore, E.E. 1999, ApJ, submitted, astro-ph/9908281

| | | | | | | | | |
|---|---|---|---|---|---|---|---|---|
| <td colspan="9" align="center">**Table 1. BATSE / BeppoSAX GRBs with Optical or Radio Counterparts**</td> |
| GRB Date | γ-ray | X-ray | γ/X | z | $D_{L_p}$ [3] | $L_p$ [4] | CCF31 Lags [5] | |
| | Peak Flux [1] | Peak Flux [2] | Ratio | | | | 0.1 | 0.5 |
| 97 05 08.90400 | 1.2 | 3.0 [6] | 25 | 0.835 [7] | 5.7 | 2.9 | 0.307 | 0.221 |
| 97 08 28.73931 | 4.9 | 1.9 [8] | 158 | 0.958 [9] | 6.7 | 16.4 | 0.028 | 0.033 |
| 97 12 14.97270 | 2.3 | 2.5 [10] | 56 | 3.412 [11] | 32.1 | 173. | 0.010 | 0.008 |
| 98 04 25.90915 | 1.1 | 2.6 [12] | 26 | 0.0085 [13] | 0.04 | $1.3\times10^{-4}$ | – | – |
| 98 05 19.51404 | 4.7 | 2.9 [14] | 100 | – | – | – | – | – |
| 98 07 03.18247 | 2.6 | 4.0 [15] | 40 | 0.967 [16] | 6.8 | 8.9 | 0.147 | 0.157 |
| 99 01 23.55535 | 16.4 | 4.0 [17] | 252 | 1.600 [18] | 12.7 | 196. | 0.015 | 0.004 |
| 99 05 10.36736 | 8.16 | 2.02 [19] | 249 | 1.619 [20] | 12.9 | 100. | 0.012 | 0.011 |

Table 1 footnotes:

[1] ph cm$^{-2}$ s$^{-1}$ (50-300 keV)

[2] $\times 10^{-8}$ erg cm$^{-2}$ s$^{-1}$ (2-10 keV)

[3] Luminosity distance in Gpc, for a universe with $(\Omega_M, \Omega_\Lambda) = (0.3, 0.7)$ and $H_o = 65$ km s$^{-1}$ Mpc$^{-1}$.

[4] Isotropic $\gamma$–ray peak luminosity in units of $10^{51}$ erg s$^{-1}$ (50–300 keV).

[5] CCF lag between BATSE channels 1 and 3 in seconds, for regions down to 0.1 and 0.5 of peak intensity, corrected for time dilation

[6] Piro, L. et al. 1997, astro-ph/9710355

[7] Metzger, M. et al. 1997, IAUC # 6655

[8] Remillard, R. et al. 1997, IAUC # 6726

[9] Djorgovski, S. 1999, ITP Conf., March 15-19, Santa Barbara, California

[10] Heise, J. et al. 1997, IAUC # 6787

[11] Kulkarni, S. et al. 1998, Nature, 393, 35

[12] Pian, E. et al. 1998, GCN # 69

[13] Tinney, C. et al. 1998 IAUC # 6896

[14] Muller, J. et al. 1998, IAUC # 6910

[15] Levine, A. et al. 1998, IAUC # 6966

[16] Djorgovski, S. et al. 1998, ApJ, 508, L17

[17] Feroci, M. et al. 1999, IAUC # 7095

[18] Djorgovski, S. et al. 1999, GCN # 251

[19] Dadina, M. et al. 1999, IAUC # 7100

[20] Vreeswijk, P. et al. 1999, GCN # 324

**Table 2. Spectral Lags for GRBs with Optical or Radio Counterparts**

| GRB Date | CCF3 (bins)[1] | CCF4 (bins)[1] | Intensity Range | Lag 31 (s)[2] cubic fit | Lag 31 (s)[2] quadratic fit | Lag 41 (s)[2] quadratic fit |
|---|---|---|---|---|---|---|
| 970508 | 28 | 36 | 0.1 | $0.564^{+0.124}_{-0.132}$ | $0.444^{+0.098}_{-0.090}$ | $0.968^{+0.212}_{-0.204}$ |
|  |  |  | 0.3 | $0.466^{+0.110}_{-0.102}$ | $0.406^{+0.088}_{-0.104}$ | $1.096^{+0.164}_{-0.242}$ |
|  |  |  | 0.5 | $0.406^{+0.114}_{-0.094}$ | $0.384^{+0.090}_{-0.084}$ | $0.922^{+0.226}_{-0.196}$ |
| 970828 | 14 | 14 | 0.1 | $0.054^{+0.012}_{-0.014}$ | $0.090^{+0.008}_{-0.026}$ | $0.110^{+0.026}_{-0.020}$ |
|  |  |  | 0.3 | $0.056^{+0.012}_{-0.016}$ | $0.092^{+0.008}_{-0.010}$ | $0.104^{+0.022}_{-0.030}$ |
|  |  |  | 0.5 | $0.064^{+0.016}_{-0.014}$ | $0.106^{+0.010}_{-0.010}$ | $0.132^{+0.028}_{-0.020}$ |
| 971214 | 10 | 30 | 0.1 | $0.042^{+0.018}_{-0.018}$ | $0.046^{+0.042}_{-0.020}$ | $0.110^{+0.458}_{-0.204}$ |
|  |  |  | 0.3 | $0.040^{+0.018}_{-0.018}$ | $0.044^{+0.048}_{-0.018}$ | $0.262^{+1.694}_{-0.310}$ |
|  |  |  | 0.5 | $0.034^{+0.024}_{-0.018}$ | $0.066^{+0.026}_{-0.048}$ | $0.132^{+0.760}_{-0.266}$ |
| 980519 | 14 | 14 | 0.1 | $0.038^{+0.016}_{-0.012}$ | $0.048^{+0.012}_{-0.010}$ | $0.110^{+0.082}_{-0.104}$ |
|  |  |  | 0.3 | $0.040^{+0.018}_{-0.014}$ | $0.050^{+0.014}_{-0.010}$ | $0.124^{+0.048}_{-0.080}$ |
|  |  |  | 0.5 | $0.044^{+0.020}_{-0.014}$ | $0.062^{+0.036}_{-0.014}$ | $0.124^{+0.056}_{-0.062}$ |
| 980703 | 34 | 42 | 0.1 | $0.290^{+0.114}_{-0.108}$ | $0.430^{+0.146}_{-0.096}$ | $0.948^{+0.288}_{-0.392}$ |
|  |  |  | 0.3 | $0.308^{+0.114}_{-0.068}$ | $0.430^{+0.138}_{-0.138}$ | $0.686^{+0.262}_{-0.292}$ |
|  |  |  | 0.5 | $0.308^{+0.110}_{-0.172}$ | $0.402^{+0.162}_{-0.134}$ | $0.756^{+0.296}_{-0.348}$ |
| 990123 | 7 | 7 | 0.1 | $0.040^{+0.012}_{-0.012}$ | $0.066^{+0.008}_{-0.008}$ | $0.074^{+0.018}_{-0.012}$ |
|  |  |  | 0.3 | $0.042^{+0.018}_{-0.014}$ | $0.062^{+0.014}_{-0.008}$ | $0.080^{+0.014}_{-0.014}$ |
|  |  |  | 0.5 | $0.010^{+0.014}_{-0.022}$ | $0.018^{+0.012}_{-0.012}$ | $0.048^{+0.014}_{-0.030}$ |
| 990510 | 7 | 7 | 0.1 | $0.032^{+0.006}_{-0.008}$ | $0.050^{+0.004}_{-0.006}$ | $0.072^{+0.046}_{-0.040}$ |
|  |  |  | 0.3 | $0.028^{+0.006}_{-0.006}$ | $0.046^{+0.006}_{-0.016}$ | $0.080^{+0.038}_{-0.016}$ |
|  |  |  | 0.5 | $0.030^{+0.008}_{-0.006}$ | $0.052^{+0.004}_{-0.002}$ | $0.090^{+0.034}_{-0.016}$ |

[1] Range of CCF fit, 64-ms bins

[2] **Uncorrected for time dilation**

Figure Captions

Fig. 1 – Pulse asymmetry / energy-shift paradigm: Solid = low energy (few × 10 keV), dashed = high energy (few × 100 keV) emission. Pulse shapes range from narrow and symmetric with negligible centroid shift with energy, to wide and asymmetric with centroid shift comparable to full width at half maximum. See Norris et al. (1996).

Fig. 2 – Time profile CCF lags between BATSE channels 1 and 4 (top) and channels 1 and 3 (bottom) are plotted for two GRB samples. Left-hand panels show gamma-ray to X-ray peak flux ratios versus CCF lags for seven GRBs in Table 1 (excluding GRB 980425). Right-hand panels show isotropic peak luminosities versus CCF lags for the subset of six GRBs with determined redshifts. The CCF lags for these six bursts have been corrected by the factor $(1+z)^{-1}$ to remove cosmic time dilation. For all panels, the symbols represent CCF lags computed above different signal levels relative to the peak: 0.1 (diamonds), 0.3 (triangles), 0.5 (squares). Individual GRBs are labeled in YYMMDD format.

Fig. 3 – Time profile CCF lags between BATSE channels 1 and 4 for 174 bright GRBs are plotted versus two hardness ratios (HRs), computed as described in text. Top (bottom) panels show HRs for channels 3 to 2 (4 to 1). Left-hand panels plot the range of CCF lags less than 1 second. Right-hand panels plot the full range of CCF lags.

Fig. 4 – Similar to Figure 3, for time profile CCF lags between BATSE channels 1 and 3 for 174 bright GRBs are plotted versus two hardness ratios.

Fig. 5 – Time profile CCF lags between BATSE channels 1 and 4 (top) and channels 1 and 3 (bottom) are plotted versus peak flux (50–300 keV) for the 174 bright GRB sample. Left and right-hand panels are as in Figure 3 and 4.

Fig. 6 – The left panel is the same as lower right panel of Figure 2, but with log-log coordinates for CCF31 lag versus peak luminosity, for the subset of six bursts with known redshifts. The dashed line is a power-law fit to the lags for intervals including count rates $> 0.1 \times$ peak intensity (square symbols), yielding $L_{53} \approx 1.3 \times (\tau/0.01 \text{ s})^{-1.14}$. In the right panel, the luminosity range is expanded to include GRB 980425 (assuming a connection with SN 1998bw), which falls below the extrapolated power-law by a factor of ~ 400–700, depending on the precise power-law slope adopted.